\documentclass[twocolumn,nofootinbib,prl,floatfix,preprintnumbers,amsmath,amssymb,groupedaddress,superscriptaddress]{revtex4}
\usepackage{dsfont}
\usepackage{graphicx} 
\usepackage{epstopdf} 
\usepackage{slashed}
\usepackage{color}
\usepackage{subfigure}
\usepackage{amssymb}

\newcommand{\GeV}{\,\mathrm{GeV}}

\newcommand{\pslash}{\!\not\! p}
\newcommand{\genericT}{{\ensuremath{\rm T}}}

\newcommand{\mptmag}{\slashed{{p}}_\genericT}

\newcommand{\figref}[1]{Figure~\ref{#1}}
\newcommand{\definmath}[2] {\def#1{\ifmmode#2\else$#2$\fi}}
\newcommand{\invfb}{\ensuremath{{\mathrm{fb}^{-1}}}}
\newcommand{\ourintlumi}{10\,\invfb{}}

\newcommand{\mtstar}{{\ensuremath{m_{\rm T}^{\star}}}}
\newcommand{\mttrue}{{\ensuremath{m_{\rm T}}}}
\newcommand{\mtbound}{{\ensuremath{m_{\rm T}^{\mathrm{bound}}}}}
\newcommand{\mtupper}{{\ensuremath{m_{\rm T}^{\mathrm{upper}}}}}
\newcommand{\mtlower}{{\ensuremath{m_{\rm T}^{\mathrm{lower}}}}}

\newcommand{\mttwo}{{\ensuremath{m_{\rm T 2}}}}
\newcommand{\ttbar}{{\ensuremath{t\bar{t}}}} 

\newtheorem{theorem}{Theorem} 

\newcommand{\qed}{\nobreak \ifvmode \relax \else
      \ifdim\lastskip<1.5em \hskip-\lastskip
      \hskip1.5em plus0em minus0.5em \fi \nobreak
      \vrule height0.75em width0.5em depth0.25em\fi}
\newcommand{\hideIt}[1]{{}}

\begin{document}   
\title{Finding heavy Higgs bosons heavier than $2m_W$ in dileptonic $W$-boson decays}
\date{\today}

\author{Alan J. Barr}
\email{a.barr@physics.ox.ac.uk}
\affiliation{Denys Wilkinson Building, Keble Road, Oxford, OX1 3RH, United Kingdom}

\author{Ben Gripaios}
\email{ben.gripaios@cern.ch}  
\affiliation{Cavendish Laboratory, Dept of Physics, JJ Thomson Avenue,
Cambridge, CB3 0HE, United Kingdom}

\author{Christopher G. Lester}
\email{lester@hep.phy.cam.ac.uk}
\affiliation{Cavendish Laboratory, Dept of Physics, JJ Thomson Avenue,
Cambridge, CB3 0HE, United Kingdom}

\begin{abstract} 
We reconsider observables for
discovering a heavy Higgs boson (with $m_h > 2m_W$) via its di-leptonic decays $h\rightarrow WW\rightarrow
\ell\nu\ell\nu$.
We show that observables generalizing the transverse mass that take
into account the fact that both of the intermediate $W$ bosons are
likely to be on-shell give a significant improvement over the variables
used in existing searches. We also comment on the application
of these observables to other decays which proceed via narrow-width intermediates.
\end{abstract}   
\maketitle 
The LHC collaborations have now excluded a heavy Higgs boson, with mass
between 127 and 600~GeV~\cite{ATLAS:2012ae,Chatrchyan:2012tx}, at the 95\% confidence level, assuming that its production and decay
proceed according to the predictions of the Standard
Model. Nevertheless,
there is a clear motivation to continue the search in
this mass region. Indeed, theories beyond the Standard Model, with
additional degrees of freedom,
typically
predict a lower cross section times branching ratio for Higgs
production and decay in any given channel, for three reasons. Firstly,
the branching ratio will necessarily be reduced whenever there are extra states into which a given Higgs boson may
decay. Secondly, in theories with an extended scalar sector, mixing
reduces the cross section to produce a given mass eigenstate. Thirdly,
since the dominant production mechanism for Higgs bosons at the LHC is
via loop-level gluon fusion, the production cross section will be
changed by the presence of additional coloured states in the
theory. If one of these states plays the r\^{o}le of cancelling the
quadratic divergence in the Higgs mass, coming from a loop of top quarks, then
it necessarily interferes destructively with the dominant contribution
to gluon fusion in the Standard Model, which also comes from a top
quark loop \cite{Low:2009di}. Furthermore, it is observed (though
sadly no theorem to the effect has been proven) that the cross-section is reduced in all
models in which the Higgs boson is composite \cite{Falkowski:2007hz,Low:2009di,Low:2010mr,Furlan:2011uq,Anastasiou:2011pi}.

One way to search for a heavy Higgs boson is to look for evidence of it decaying
to the fully leptonic final state $l \nu l \nu$ (where $l$ is an
electron or muon) which results from an
intermediate pair of (possibly virtual) $W$-bosons
\cite{Glover:1988fn,Barger:1990mn,Dittmar:1996ss}.
 A key step in many such analyses comprises a
selection, or cut, based on a variable that should have a strong
correlation with the mass of the Higgs boson in `signal' events.
Typically this variable will be a transverse mass of some
kind, of which many have been described in the literature \cite{Barr:2010zj,Barr:2011xt}.
The transverse mass variable $\mttrue$ of Ref.~\cite{Barr:2009mx} (see also \cite{Barger:1987re,Glover:1988fn})
is used in current LHC $h\rightarrow W W\rightarrow l\nu l\nu$
searches.\footnote{The variable $\mttrue$ was referred to
  as $m_{\rm T}^{\mathrm{true}}$ in
  \cite{Barr:2009mx} to distinguish it from other, approximate transverse
  mass variables in use at the
  time; we now refer to it simply as $\mttrue$, in line with both the ATLAS
  and CMS
  collaborations.}

It has previously been observed~\cite{Barr:2011he} 
that a kinematic variable \mtbound{} (which, for reasons that will
become clear, we refer to henceforth as $\mtlower$) which demands intermediate mass-shell constraints
for the tau leptons might prove advantageous in
$h\rightarrow \tau\tau$ decays.
As was indicated in that paper, the same variable 
(after the trivial replacement $m_\tau\rightarrow m_W$) 
might also prove useful in the context of  $h\rightarrow WW$
when a heavy Higgs boson has mass greater than $2m_W$,
since the intermediate $W$-bosons should be produced close 
to their mass shells.
Substantiation of this suggestion was left open to further study and
one of our intentions here is to perform that follow-up. We
  also intend to go somewhat further, in providing a complete discussion
of the variables that characterize the kinematics of $h\rightarrow WW
\rightarrow l\nu l \nu $ decays, in a sense that will be made explicit
below. We will find that the pre-existing variables
$\mttrue$ or $\mtstar$ \cite{Barr:2011ux} provide a complete description in cases
where neither or either, respectively, of the intermediate $W$-bosons
is produced on-shell.\footnote{In the mass range $m_W < m_h < 2 m_W$, one might reasonably
assume that one or other of the $W$ bosons is produced on shell
\cite{Djouadi:2005gi}.
However, simulations reported in \cite{Barr:2011ux} found no significant
improvement in discovery potential when $\mtstar$ was
substituted for $\mttrue$ in existing ATLAS collaboration searches. 
}
 A complete description of the cases when both of
the $W$-bosons
are produced on-shell requires not only the extant variables
$\mtlower$ and $\mttwo$ \cite{Lester:1999tx,Barr:2003rg},
but also a new variable, which we denote $\mtupper$. Our simulations suggest, moreover, that $\mtupper$
gives a significant improvement in discovery potential for $h\rightarrow WW
\rightarrow l\nu l \nu $. 

Having finished our preamble, we now describe more explicitly how
variables of this kind may provide a complete characterization of the
kinematics of events. Since the decay of the parent Higgs boson involves
invisible neutrinos, one cannot find a variable that measures its mass
directly; rather the best one can do is to find variables that bound
the mass of the parent in some way. Myriad variables of this type may
exist
and what one would like to do, presumably, is to
find the variable or variables that give the {\em optimal} bounds on
the mass in an event.

This search for the optimal variables can be done in a more-or-less definitive
  way if one is willing to restrict one's attention to kinematics
  alone. That is to say, suppose one assumes that a signal event
  corresponds to some decay topology
 (here the
  decay of a parent Higgs into two intermediate $W$ 
  resonances, followed by decays of each intermediate into a
  combination of visible and invisible daughter particles). One may
  then write down the various kinematic constraints (corresponding to
  energy-momentum conservation and the mass-shell conditions) and ask
  which values of the {\em a priori} unknown masses are allowed, in
  the sense that the kinematic constraints then admit solutions in
  which the unmeasured momenta and energies in the event are real and
  real-positive, respectively. The resulting allowed mass region is an
  event observable and encodes all of the information which can be
  gleaned about the masses in an event using kinematics alone. The
  boundary of the allowed region can be described in terms of one or
  more relations between the masses, which are themselves event
  observables and which give the optimal bounds on the mass
    in an event. Given multiple events, the allowed mass region is
    obtained as the intersection of the allowed regions for each event
    and is itself an observable.\footnote{Whether this observable
    is the optimal observable for discovering the parent is
      unclear. To settle this, one would first have to define what one
    meant by an optimal discovery observable. For discussion, see {\em e.g.} \cite{Kim:2009si,Rujula:2011qn,DeRujula:2012ns}.}

This abstract recipe leads, in many cases, to simple, pre-existing
observables. In the canonical example of a decay of a parent to a
single invisible daughter (together with possibly multiple visible particles), the observable that
results \cite{Barr:2009jv} is the familiar transverse mass, $m_T$ \cite{vanNeerven:1982mz}, invoked in the discovery of the $W$-boson
\cite{Arnison:1983rp} in its decay to a charged lepton and a
neutrino; for single parent decays into multiple invisibles, the
resulting variable \cite{Barr:2009mx} is $\mttrue$; for identical decays of a
pair of 
parents
into invisible particles,
the variable \cite{Serna:2008zk,Cheng:2008hk,Barr:2009jv} is $\mttwo$.

Thus, these variables encode all of the information that is available
from kinematic considerations alone (subject to an assumed decay
topology) and there is no point
in trying to devise further variables to glean further information from kinematics.
\footnote{The equivalence between variables and kinematics is
also important for the purposes of determining which particle masses can be
measured using kinematics alone: as an example, for a (possibly pair-produced) parent particle that decays off-
shell to multiple invisible particles, it is a theorem that one can
-- at least in principle -- 
measure the parent mass and the sum of the masses of the invisible
daughter particles using kinematic information alone
\cite{Cho:2007qv,Gripaios:2007is,Barr:2007hy,Cho:2007dh}.}

In the next Section, we firstly try to prove the equivalence
between the kinematic constraints with one intermediate on-shell and the variable $\mtstar$
 and secondly the equivalence between the constraints with both
 intermediates on-shell and the variables
$\mtlower$ and $\mttwo$.
Our first
attempt at a proof succeeds: $\mtstar$ is equivalent to all of the
information contained in the kinematic constraints when one
intermediate is on-shell. But for the
topology with both intermediates on shell, we fail. Indeed, we show
that $\mtlower$ and $\mttwo$ do not capture all of the information in the
kinematic constraints. Rather there is a third, distinct variable, which we
call $\mtupper$, and which gives an {\em upper} bound on the mass of
the Higgs (or whatever parent particle is being
considered).

In what follows we show that while $\mtlower$ only marginally
enhances the discovery potential for a Higgs boson using current ATLAS
collaboration search strategies, $\mtupper$ gives a
significant improvement.
In an Appendix, we discuss issues related to the existence of
$\mtlower$ and $\mtupper$.
\section{Kinematic Equivalence \label{sec:kin}}
Since we are concerned with the three transverse masses $\mttrue$,
$\mtstar$ and $\mtlower$, it is helpful to recall and compare their
definitions. All three make the assumption that the observed $l \nu l
\nu$ final state resulted from the decay of a single progenitor
particle or resonance (i.e.~the Higgs in the case of the signal) whose
mass is, {\em a priori}, unknown. All three assert that the sum of the transverse momenta
of the neutrinos should equal the observed missing transverse
momentum.  All three accept that: (i) the way the transverse momentum
is shared between the two neutrinos is unknown, and (ii) that the
longitudinal momentum component of each neutrino is unknown.  Finally,
all three transverse masses are defined to be {\em the greatest
possible lower bound for the unknown mass of that progenitor
particle}, subject to consistency and any remaining
constraints.\footnote{Equivalently, they may be defined to be
the smallest possible progenitor mass obtainable by a search over all
unknown neutrino momentum components that satisfy the remaining
constraints.  Indeed, this brute force search technique is used in the
authors' implementations of $\mtstar$ and $\mtlower$ calculators.
These {\tt C++} implementations, free of external dependencies, are
available on request.  A simple analytic answer is available for
$\mttrue$ -- see Eq.~(1) of \cite{Barr:2011ux}.}  It is only in
these remaining constraints that the transverse masses differ:
$\mttrue$ applies none, $\mtstar$ permits only neutrino momenta that
place {\em at least one} $W$-boson on mass-shell, while $\mtlower$
allows only neutrino momenta that place {\em both} $W$-bosons on
mass-shell. In short, history would have been kinder to us
had $\mttrue$, $\mtstar$ and $\mtlower$ been named $m_T^{W^\star
W^\star}$, $m_T^{W W^\star}$, and $m_T^{WW}$ respectively.  The variables and constraints are summarized in Table~\ref{tab:constraints}. 


\begin{table*}\renewcommand\arraystretch{1.5}
\centering
\begin{minipage}[c]{0.4\linewidth}
\centering
\begin{tabular}{l l l l}
\hline
Name       & Ref. & \begin{minipage}[l]{1.5cm}Bounding\end{minipage} & Constraints \\
\hline
$\mttrue$  & \cite{Barger:1987re,Glover:1988fn,Barr:2009mx} & $H^\mu H_\mu$ & 
\eqref{eq:constraints_all_begin} to \eqref{eq:constraints_all_end} \\
$\mtstar$  & \cite{Barr:2011ux} & $H^\mu H_\mu$ & 
\eqref{eq:constraints_all_begin} to \eqref{eq:constraints_all_end} and either \eqref{eq:constraints_tau1} or \eqref{eq:constraints_tau2}\\
$\mtlower$ & \cite{Barr:2011he} & $H^\mu H_\mu$ & 
\eqref{eq:constraints_all_begin} to \eqref{eq:constraints_tau2} \\
\hline
\end{tabular}
\end{minipage}\hspace{5mm}
\begin{minipage}[c]{0.4\linewidth}
\centering
\begin{eqnarray}\label{eq:constraints_all_begin}p_\nu^{\mu} p_{\nu\mu}
  = p_l^{\mu} p_{l\mu} &=& 0,  \\
p_\nu^{\prime \mu} p_{\nu\mu}^\prime = p_l^{\prime \mu}
p_{l\mu}^\prime &=& 0, \\
p_{\nu T}+p_{\nu T}^\prime &=& \slashed{p}_T, \label{eq:constraints_all_end}\\
(p_\nu^\mu+p_l^\mu)(p_{\nu\mu}+p_{l\mu}) &=& m_W^2,  \label{eq:constraints_tau1}\\
(p_\nu^{\prime \mu}+p_l^{\prime \mu})(p_{\nu}^{\prime \mu}+p_{l\mu}^\prime) &=& m_W^2.  \label{eq:constraints_tau2}
\end{eqnarray}
\end{minipage}
\vskip 5mm
\begin{minipage}[c]{0.4\linewidth}{\centering \bf (a)} Variables\end{minipage}
\hspace{5mm}
\begin{minipage}[c]{0.4\linewidth}{\centering \bf (b)} Constraints\end{minipage}
\caption{\label{tab:constraints}
A comparison of the three mass-bound variables. The final state
four-momenta of the two visible leptons are $p_l^{\mu}$ and
$p_l^{\prime \mu}$ respectively. 
The hypothesized four-momenta of the two neutrinos are given by the two $p_\nu^\mu$. The missing transverse momentum is $\slashed{p}_T$.
In all three cases, the variable named in {\bf (a)} is defined to be
the maximal lower bound on the invariant $H^\mu H_\mu$, where $H^\mu =
p_\nu^{\mu}+p_l^{\mu} + p_\nu^{\prime \mu}+p_l^{\prime \mu}$, subject to the indicated subset of the constraints shown in {\bf (b)}. 
}
\end{table*}
We now attempt to prove the equivalence between the variables
$\mtstar$ and $\mtlower$ and the corresponding kinematic constraints.
\subsection{One intermediate on-shell}
The theorem to be proved is a formalization of the colloquial: $\mtstar$
is equivalent to all the information contained in the kinematic
constraints. To wit, in a notation in which $p^\mu =
(E,p_T,q)$:
\begin{theorem}
The kinematic constraints \eqref{eq:constraints_all_begin} to
\eqref{eq:constraints_all_end} and either \eqref{eq:constraints_tau1} or \eqref{eq:constraints_tau2}
admit a solution with momenta in $\mathbb{R}$ and energies in
$\mathbb{R}^+$ iff.\ 
\begin{gather}
m_h \geq \mtstar.
\end{gather}
\end{theorem}
Here, $\mtstar$ is defined as the minimum value of 
\begin{gather}
m_h \geq \mtstar 
\end{gather}
obtained by varying the unobserved momenta in $\mathbb{R}$ and energies in
$\mathbb{R}^+$. This definition makes the necessity of $m_h \geq
\mtstar $ obvious; it is the sufficiency of the condition that
requires deliberation.

So, does $m_h \geq \mtstar $ imply that the kinematic constraints have
a solution? One way to show this would be to consider every $m_h \geq
\mtstar $ and explicitly construct a solution therefor. An easier way
is to show that one can find solutions corresponding to arbitrarily
large $m_h$ and then to invoke continuity and the intermediate value
theorem to show that one can find solutions for any $m_h$ between
$\mtstar$ and $\infty$.

Let us then exhibit a solution for arbitrarily large $m_h$ that
satisfies constraints \eqref{eq:constraints_all_begin} to
\eqref{eq:constraints_tau1} (if the other $W$ is on-shell, the
required solution can be obtained by interchanging primed and unprimed
quantities). It is
given by
\begin{align}
p_\nu^{\prime \mu} &= \lim_{q_\nu^{\prime} \rightarrow \infty } (\sqrt{q_\nu^{\prime2} + \slashed{p}_T^2},\slashed{p}_T, q_\nu^\prime ), \\
p_\nu^{\mu} &= (|q_\nu |, 0, q_\nu) 
\end{align}
where
$q_\nu$ is given by either of 
\begin{gather}
q_\nu = \frac{m_W^2 }{2p_{lT}^2} \left(q_l \pm \sqrt{q_l^2 +p_{lT}^2 }\right).
\end{gather}
This yields a value of $m_h$ given by
\begin{align}
m_h^2 &= \lim_{q_\nu^{\prime} \rightarrow \infty } (m_W^2 + 2(p_l + p_\nu) \cdot
p_l^\prime + 2(p_l^\prime  + p_l + p_\nu  ) \cdot p_\nu^\prime)  \nonumber\\
&\rightarrow 2((E_l^\prime+ E_l + |q_\nu | )
\sqrt{q_\nu^{\prime2} + \slashed{p}_T^2}\nonumber\\
& - (p_{lT}^\prime
+ p_{lT})
\cdot \slashed{p}_T  
 -  (q_l^\prime + q_l + q_\nu
)q_\nu^{\prime})) \nonumber\\
&\rightarrow 2((E_l^\prime - q_l^\prime)  + (E_l - q_l) + (|q_\nu | -q_\nu)) q_\nu^\prime,
\end{align}
where in the last line we assumed $q_\nu^\prime > 0$ without loss of
generality (this amounts to a choice of which proton beam constitutes
the $+z$ direction). The terms in parentheses in the coefficient are each positive
semi-definite and so no cancellation of the coefficient of $q_\nu^\prime$ can
occur. Thus $m_h$ grows without bound unless all terms vanish. To do
so, both leptons must have vanishing transverse
momenta, in which case they would fall outside the detector acceptance
and this would not be identified as a di-lepton plus missing momentum event.\qed 
\subsection{Both intermediates on-shell}
It is now easy to show that an analogous theorem cannot be proven for
$\mtlower$,
which applies to the topology in which 
both constraints \eqref{eq:constraints_tau1} and \eqref{eq:constraints_tau2}
are satisfied. The reason is that one cannot find
solutions to the kinematic constraints corresponding to arbitrarily
large $m_h$. Indeed, one can easily convince oneself that to get
arbitrarily large $m_h$, at least one of the neutrino momenta must
become arbitrarily large. But each neutrino is now produced in the
decay of a $W$-boson, which simultaneously results in a charged lepton
of finite (and measured) momentum. Now, there is only one way in which
a two-body decay can produced one daughter of finite momentum and
another with infinite momentum in the lab frame: the rest frame of the
$W$ must be infinitely boosted exactly in the direction of motion of the
neutrino (as measured in the $W$ rest frame). This doesn't work,
because the lepton momenta are then not only finite but arbitrarily
small, in contrast with their measured values in a generic event.

Thus, there is more information in the kinematic constraints than is
captured by $\mtlower$
alone. In particular, since $m_h$ cannot reach
arbitrarily large values in events, there is a distinct variable,
obtained by {\em maximizing } the sum of lepton and neutrino momenta
in events, subject to the above constraints. This variable will be
bounded {\em below} by $m_h$ and we call it $\mtupper$. By the intermediate value theorem, this
variable, together with $\mtlower$ does contain all of the information
in kinematics.
\section{Monte Carlo Simulations}
Simulations, similar to those performed in Ref~\cite{Barr:2009mx}, are used to test 
the extent to which using $\mtlower$ and $\mtupper$ might be expected to enhance
the statistical significance of a $h\rightarrow WW$ signal above the dominant 
$WW$ continuum background and the \ttbar{} background.
We use the {\tt HERWIG}~6.505~\cite{Corcella:2002jc,Marchesini:1991ch} Monte
Carlo generator, with LHC beam conditions ($\sqrt{s}=7~\rm{TeV}$).
Our version of the generator includes the fix to the $h \to WW^{(*)}$
spin correlations described in \cite{herwig-bug-fix}.
We generate unweighted events for Standard Model Higgs boson production ($gg\rightarrow h$)
and for the dominant backgrounds, $q \bar{q} \rightarrow WW$, and \ttbar{} production.\footnote{Other backgrounds, such as $Z\rightarrow \tau\tau$, are rendered sub-dominant by the cuts discussed below.}
We use the leading order Standard Model cross section for all values of $m_h$.

The missing transverse momentum is calculated from the negative sum of the $p_T$ of visible particles
within the fiducial region $p_T>0.5$ GeV and $|\eta|<5$.
The detector resolution is simulated by smearing the magnitude of the missing momentum vector 
with a Gaussian resolution function of width
$\sigma_{\mptmag}/\mptmag = 0.4\,{\rm GeV}^{1/2}/\sqrt{\Sigma}$
where $\Sigma$ is the sum of the $|\vec{p}_{\genericT}|$ of all visible fiducial particles.

For each value of $m_h$, fifty pseudo-experiments are performed,
each corresponding to \ourintlumi.
Selection cuts are applied, requiring:
\begin{itemize}
\item{Exactly two leptons $\ell\in\{e,\mu\}$ with $p_T > 15$~GeV and $|\eta|<2.5$}
\item{Missing transverse momentum, $\pslash_T > 30$~GeV}
\item{ $12~\GeV < m_{\ell\ell} < 300~\GeV$}
\item{No jet with $p_T > 20$~GeV }
\item{$Z\rightarrow \tau\tau$ rejection: the event was rejected if
  $|m_{\tau\tau}-m_{Z}|<25~\GeV$ and
  $0<x_i<1$ for both $i\in \{1,2\}$\footnote{The variable $x_i$ is the
  momentum fraction of the $i$th tau carried by its daughter lepton and $m_{\tau\tau}$ is the di-tau invariant mass.
  They are calculated using the approximation that each $\tau$ was collinear with its daughter lepton.}}
\end{itemize}

For each of the three candidate discriminant variables \mttrue{}, \mtlower{}, and \mtupper{}, and for each trial value of $m_h$, a likelihood is calculated for each of two hypotheses $\mathcal{H}$: a background-only ($B$) hypothesis and a signal plus background hypothesis ($S+B$).
The log likelihood 
\[
\log \mathcal{L}_{\mathcal{H}} = \left< \log \prod_{i\in\mathcal{D}} P\left(n_i^{(j)}|x_i^\mathcal{H}\right)\right>_{j=1\ldots50}
\]
is calculated from the average over 50 pseudo-experiments. For each pseudo-experiment $j$ the number of events observed in the $i$th bin is $n_i^{(j)}$. The expected number of events $x_i^\mathcal{H}$ for each bin and for each of the two hypotheses $\mathcal{H}$ is calculated from independent Monte Carlo samples containing $10^6$ events for each trial value of $m_h$ and $10^7$ events for the $WW$ background.
The likelihood for each $n_i^{(j)}$ is calculated using a Poisson distribution with mean $x_i^\mathcal{H}$.
The difference $-2\Delta \log \mathcal{L} = -2\left( \log\mathcal{L}_{S+B} - \log\mathcal{L}_B \right)$
quantifies the statistical significance with which we can expect the Higgs boson signal to be observed.

\begin{figure}
 \begin{center}
\includegraphics[width=0.99\columnwidth]{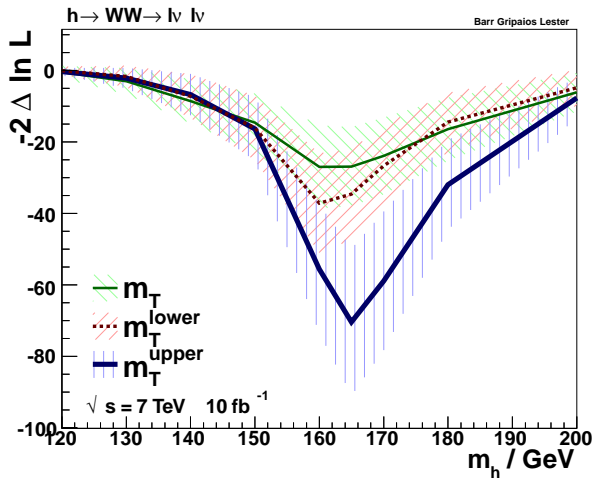}
\caption{
Higgs boson discovery potential as a function of $m_h$, for each of three different kinematical variables, \mttrue, \mtlower, and \mtupper. 
The centre of each band indicates the difference in log likelihood 
between models with and without a Higgs boson contribution.
Lower values correspond to better discovery potential.
The half-width of each band gives the root-mean-squared variation over the 50 pseudo-experiments. 
The integrated luminosity simulated is \ourintlumi.
}
\label{fig:wasp}
\end{center}
\end{figure}

\begin{figure}
 \begin{center}
\includegraphics[width=0.99\columnwidth]{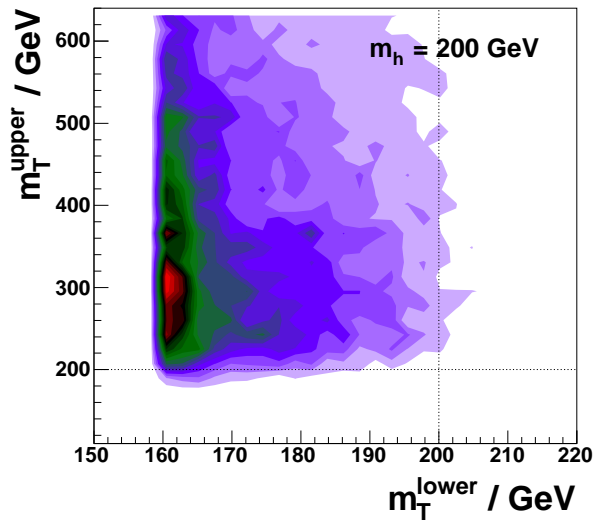}
\caption{
The distribution of \mtupper{} vs. \mtlower{} for a Higgs boson signal sample with $m_h=200$~GeV.
}
\label{fig:2d}
\end{center}
\end{figure}

The discovery potential using separately \mttrue, \mtlower, or \mtupper{}, 
is shown in Figure~\ref{fig:wasp}. 
The first result is that when $m_h\approx 2 m_W$ the doubly-on-shell lower bound variable, \mtlower{} has significantly better sensitivity than the simple transverse mass \mttrue{} (or indeed the singly-on-shell variable \mtstar{}, which is almost indistinguishable from \mttrue~\cite{Barr:2011ux}).
Secondly, we observe that the distribution of the doubly-constrained {\em upper} bound \mtupper{} shows a markedly greater discovery potential again, relative to either \mttrue{} or \mtlower{}.
Since any event satisfying the constraints of Table~\ref{tab:constraints} will generate both a lower and an upper bound, and since those bounds are not closely correlated (\figref{fig:2d}) each adds information relative to the other. 

\section{Other final states}
We note that it is also possible to construct equivalent bounds for
any other decays of the form $A
\rightarrow B + C$ where $B$ and $C$ are produced near their mass shell and each decays to an invisible particle and one or more visible particles. A case relevant to LHC Higgs physics is the decay mode $h\rightarrow \tau\tau$, followed by semi-invisible decay of each of the $\tau$ leptons. Previous simulations~\cite{Barr:2011he} indicated that adding the tau mass constraints led to improved discrimination of the signal from the dominant $Z\rightarrow \tau\tau$ background, using the lower bound \mtlower~\cite{Barr:2011he}. However further simulations indicate that there seems to be little additional discrimination available from the upper bound in this case, perhaps because of the large hierarchy in masses between $m_h$ and $m_\tau$.

\section{Conclusion}

When a heavy higgs boson with mass $>2 m_W$ decays via $h\rightarrow WW\rightarrow \ell\nu\ell\nu$,
we have shown that one can use knowledge about the narrow width of the intermediate $W$ bosons
to construct both upper and lower bounds on $m_h$. 
We demonstrate that these two bounds together make maximal use of the kinematic information about $m_h$ in each event.
Since the two bounds are reasonably uncorrelated, the ATLAS and CMS experimental collaborations should be able to obtain improved kinematic sensitivity in an analysis which simultaneously makes use of both bounds.

We note that the same suite of variables can find use for discovery or mass determination
in other decays of the form $A\rightarrow B + C $ followed by semi-invisible decay of each narrow-width intermediate particle.

\begin{acknowledgments}
This work was supported by the Science and Technology Facilities Council of the United Kingdom, by the Royal Society, by the Institute for Particle Physics Phenomenology, by Merton College, Oxford, and by Peterhouse, Cambridge.
We gratefully acknowledge T.J.~Khoo for helpful comments.
\end{acknowledgments}
\appendix
\section{Appendix: Remarks on the existence of  $\mtlower$ and  $\mtupper$}
Note that there exist `input momenta' for which it is not possible to
satisfy all the constraints required by $\mtlower$ (and $\mtupper$, to
which the following arguments equally apply).  For events containing
such momenta, a value of $\mtlower$ cannot be computed.  In effect,
the variable does not exist as those events are incompatible with the
assumptions at face value.\footnote{Note also that the variables $\mttrue$ and $\mtstar$ exist for all input momenta, and are therefore usable even when $\mtlower$ is not.  See discussion in \cite{Barr:2011he}.}
It is known \cite{Barr:2011he} that $\mtlower$ exists if and only if $m_{T2}$ \cite{Lester:1999tx,Barr:2003rg} for the dilepton system is less than $m_W$. This condition may easily be violated if one or both $W$-bosons is off mass-shell, confirming that there is little sense in applying $\mtlower$ to events with $m_h < 2 m_W$.  The criterion $m_{T2} < m_W$ can also be violated in  $m_h > 2 m_W$ events with sufficiently large detector mis-measurements, even if both $W$-bosons are produced on-shell.  The largest source of mis-measurement is the missing transverse momentum, whose distribution can have
a large, non-Gaussian tail. Though not discussed in
\cite{Barr:2011he}, this potential problem might be thought more worrisome in the
case of $h \rightarrow \tau \tau$ decays, since the typical size of an
LHC $\slashed{p}_T$ mismeasurement greatly exceeds $m_\tau$. It is thus
hard to see, na\"{\i}vely, how the condition $m_{T2} < m_\tau$ may be
maintained. Nonetheless, one is assisted by a property of the 
$m_{T2}$ variable under which large mis-measurements of
$\slashed{p}_T$ can lead, in a non-negligible number of events, to
small changes (relative to $m_\tau$) in the value of $m_{T2}$,
provided that the intermediate particles in the production process are
sufficiently light. Indeed, in the limit that the intermediate pair
  of particles are massless, their decay products are collinear and so
  the true $\slashed{p}_T$ vector lies in the smaller of the two sectors bounded by the two visible transverse momenta. It follows that for such
configurations $m_{T2}$ must vanish, as there exists a splitting of the $\slashed{p}_T$ (in fact the `true' splitting) for which the transverse mass of each pair of decays is zero.  That $m_{T2}$ is `small' for well-measured events in this limit is not surprising -- after all, $m_{T2}$ is supposed to be bounded above by the mass of either member of the pair, and we are taking the limit in which those masses go to zero. What is surprising, if anything, is that $m_{T2}$ can be forced {\em exactly} to zero in a large number of events, even for finite yet small intermediate particle masses.  In this sense we can say that $m_{T2}$ goes to zero faster than $m_{\tau}$.  It follows that for such signal events, any mis-measurement of $\slashed{p}_T$ (even large ones) that keeps it `between' the visible particles' transverse momenta, leads to no observable change in $m_{T2}$. 
Moving away from the above limit, such that the intermediate particles now have non-zero but still small intermediate masses, we see that changes in $m_{T2}$ can still be small, even
for large mis-measurements of $\slashed{p}_T$, provided that
mis-measured $\slashed{p}_T$ remains in the appropriate region of the
transverse plane. 

This goes some way towards explaining why it was found in
\cite{Barr:2011he} that $\mtlower$ could remain well-defined for a
large number of $h \rightarrow \tau \tau$ events, even in the presence of momentum mis-measurements. For $h \rightarrow WW$ events this issue is much less of
a concern; we find that $\mtlower$ exists for all but a few {\em per cent} of the $h \rightarrow WW$ signal events we simulate.  We attribute the greater frequency with which $\mtlower$ exists to the greater ease with which the momentum mis-measurement errors (which have a natural scale far below $m_W$) may be incorporated into the intermediate particle mass constraints (whose natural scale is $m_W$). 
\bibliography{Heavy}

\begin{thebibliography}{34}
\expandafter\ifx\csname natexlab\endcsname\relax\def\natexlab#1{#1}\fi
\expandafter\ifx\csname bibnamefont\endcsname\relax
  \def\bibnamefont#1{#1}\fi
\expandafter\ifx\csname bibfnamefont\endcsname\relax
  \def\bibfnamefont#1{#1}\fi
\expandafter\ifx\csname citenamefont\endcsname\relax
  \def\citenamefont#1{#1}\fi
\expandafter\ifx\csname url\endcsname\relax
  \def\url#1{\texttt{#1}}\fi
\expandafter\ifx\csname urlprefix\endcsname\relax\def\urlprefix{URL }\fi
\providecommand{\bibinfo}[2]{#2}
\providecommand{\eprint}[2][]{\url{#2}}

\bibitem[{\citenamefont{Aad et~al.}(2012)}]{ATLAS:2012ae}
\bibinfo{author}{\bibfnamefont{G.}~\bibnamefont{Aad}} \bibnamefont{et~al.}
  (\bibinfo{collaboration}{ATLAS Collaboration}), \bibinfo{journal}{Phys.Lett.}
  \textbf{\bibinfo{volume}{B710}}, \bibinfo{pages}{49} (\bibinfo{year}{2012}),
  \eprint{1202.1408}.

\bibitem[{\citenamefont{Chatrchyan et~al.}(2012)}]{Chatrchyan:2012tx}
\bibinfo{author}{\bibfnamefont{S.}~\bibnamefont{Chatrchyan}}
  \bibnamefont{et~al.} (\bibinfo{collaboration}{CMS Collaboration})
  (\bibinfo{year}{2012}), \eprint{1202.1488}.

\bibitem[{\citenamefont{Low et~al.}(2010)\citenamefont{Low, Rattazzi, and
  Vichi}}]{Low:2009di}
\bibinfo{author}{\bibfnamefont{I.}~\bibnamefont{Low}},
  \bibinfo{author}{\bibfnamefont{R.}~\bibnamefont{Rattazzi}}, \bibnamefont{and}
  \bibinfo{author}{\bibfnamefont{A.}~\bibnamefont{Vichi}},
  \bibinfo{journal}{JHEP} \textbf{\bibinfo{volume}{04}}, \bibinfo{pages}{126}
  (\bibinfo{year}{2010}), \eprint{0907.5413}.

\bibitem[{\citenamefont{Falkowski}(2008)}]{Falkowski:2007hz}
\bibinfo{author}{\bibfnamefont{A.}~\bibnamefont{Falkowski}},
  \bibinfo{journal}{Phys. Rev.} \textbf{\bibinfo{volume}{D77}},
  \bibinfo{pages}{055018} (\bibinfo{year}{2008}), \eprint{0711.0828}.

\bibitem[{\citenamefont{Low and Vichi}(2010)}]{Low:2010mr}
\bibinfo{author}{\bibfnamefont{I.}~\bibnamefont{Low}} \bibnamefont{and}
  \bibinfo{author}{\bibfnamefont{A.}~\bibnamefont{Vichi}}
  (\bibinfo{year}{2010}), \eprint{1010.2753}.

\bibitem[{\citenamefont{Furlan}(2011)}]{Furlan:2011uq}
\bibinfo{author}{\bibfnamefont{E.}~\bibnamefont{Furlan}}
  (\bibinfo{year}{2011}), \eprint{1106.4024}.

\bibitem[{\citenamefont{Anastasiou et~al.}(2011)\citenamefont{Anastasiou,
  Buhler, Herzog, and Lazopoulos}}]{Anastasiou:2011pi}
\bibinfo{author}{\bibfnamefont{C.}~\bibnamefont{Anastasiou}},
  \bibinfo{author}{\bibfnamefont{S.}~\bibnamefont{Buhler}},
  \bibinfo{author}{\bibfnamefont{F.}~\bibnamefont{Herzog}}, \bibnamefont{and}
  \bibinfo{author}{\bibfnamefont{A.}~\bibnamefont{Lazopoulos}}
  (\bibinfo{year}{2011}), \eprint{1107.0683}.

\bibitem[{\citenamefont{Glover et~al.}(1988)\citenamefont{Glover, Ohnemus, and
  Willenbrock}}]{Glover:1988fn}
\bibinfo{author}{\bibfnamefont{E.~W.~N.} \bibnamefont{Glover}},
  \bibinfo{author}{\bibfnamefont{J.}~\bibnamefont{Ohnemus}}, \bibnamefont{and}
  \bibinfo{author}{\bibfnamefont{S.~S.~D.} \bibnamefont{Willenbrock}},
  \bibinfo{journal}{Phys. Rev.} \textbf{\bibinfo{volume}{D37}},
  \bibinfo{pages}{3193} (\bibinfo{year}{1988}).

\bibitem[{\citenamefont{Barger et~al.}(1991)\citenamefont{Barger, Bhattacharya,
  Han, and Kniehl}}]{Barger:1990mn}
\bibinfo{author}{\bibfnamefont{V.~D.} \bibnamefont{Barger}},
  \bibinfo{author}{\bibfnamefont{G.}~\bibnamefont{Bhattacharya}},
  \bibinfo{author}{\bibfnamefont{T.}~\bibnamefont{Han}}, \bibnamefont{and}
  \bibinfo{author}{\bibfnamefont{B.~A.} \bibnamefont{Kniehl}},
  \bibinfo{journal}{Phys.Rev.} \textbf{\bibinfo{volume}{D43}},
  \bibinfo{pages}{779} (\bibinfo{year}{1991}).

\bibitem[{\citenamefont{Dittmar and Dreiner}(1997)}]{Dittmar:1996ss}
\bibinfo{author}{\bibfnamefont{M.}~\bibnamefont{Dittmar}} \bibnamefont{and}
  \bibinfo{author}{\bibfnamefont{H.~K.} \bibnamefont{Dreiner}},
  \bibinfo{journal}{Phys.Rev.} \textbf{\bibinfo{volume}{D55}},
  \bibinfo{pages}{167} (\bibinfo{year}{1997}), \eprint{hep-ph/9608317}.

\bibitem[{\citenamefont{Barr and Lester}(2010)}]{Barr:2010zj}
\bibinfo{author}{\bibfnamefont{A.~J.} \bibnamefont{Barr}} \bibnamefont{and}
  \bibinfo{author}{\bibfnamefont{C.~G.} \bibnamefont{Lester}},
  \bibinfo{journal}{J. Phys.} \textbf{\bibinfo{volume}{G37}},
  \bibinfo{pages}{123001} (\bibinfo{year}{2010}), \eprint{1004.2732}.

\bibitem[{\citenamefont{Barr et~al.}(2011{\natexlab{a}})}]{Barr:2011xt}
\bibinfo{author}{\bibfnamefont{A.~J.} \bibnamefont{Barr}} \bibnamefont{et~al.}
  (\bibinfo{year}{2011}{\natexlab{a}}), \eprint{1105.2977}.

\bibitem[{\citenamefont{Barr et~al.}(2009{\natexlab{a}})\citenamefont{Barr,
  Gripaios, and Lester}}]{Barr:2009mx}
\bibinfo{author}{\bibfnamefont{A.~J.} \bibnamefont{Barr}},
  \bibinfo{author}{\bibfnamefont{B.}~\bibnamefont{Gripaios}}, \bibnamefont{and}
  \bibinfo{author}{\bibfnamefont{C.~G.} \bibnamefont{Lester}},
  \bibinfo{journal}{JHEP} \textbf{\bibinfo{volume}{07}}, \bibinfo{pages}{072}
  (\bibinfo{year}{2009}{\natexlab{a}}), \eprint{0902.4864}.

\bibitem[{\citenamefont{Barger et~al.}(1988)\citenamefont{Barger, Han, and
  Ohnemus}}]{Barger:1987re}
\bibinfo{author}{\bibfnamefont{V.~D.} \bibnamefont{Barger}},
  \bibinfo{author}{\bibfnamefont{T.}~\bibnamefont{Han}}, \bibnamefont{and}
  \bibinfo{author}{\bibfnamefont{J.}~\bibnamefont{Ohnemus}},
  \bibinfo{journal}{Phys. Rev.} \textbf{\bibinfo{volume}{D37}},
  \bibinfo{pages}{1174} (\bibinfo{year}{1988}).

\bibitem[{\citenamefont{Barr et~al.}(2011{\natexlab{b}})\citenamefont{Barr,
  French, Frost, and Lester}}]{Barr:2011he}
\bibinfo{author}{\bibfnamefont{A.~J.} \bibnamefont{Barr}},
  \bibinfo{author}{\bibfnamefont{S.~T.} \bibnamefont{French}},
  \bibinfo{author}{\bibfnamefont{J.~A.} \bibnamefont{Frost}}, \bibnamefont{and}
  \bibinfo{author}{\bibfnamefont{C.~G.} \bibnamefont{Lester}}
  (\bibinfo{year}{2011}{\natexlab{b}}), \eprint{1106.2322}.

\bibitem[{\citenamefont{Barr et~al.}(2011{\natexlab{c}})\citenamefont{Barr,
  Gripaios, and Lester}}]{Barr:2011ux}
\bibinfo{author}{\bibfnamefont{A.~J.} \bibnamefont{Barr}},
  \bibinfo{author}{\bibfnamefont{B.}~\bibnamefont{Gripaios}}, \bibnamefont{and}
  \bibinfo{author}{\bibfnamefont{C.~G.} \bibnamefont{Lester}}
  (\bibinfo{year}{2011}{\natexlab{c}}), \eprint{1108.3468}.

\bibitem[{\citenamefont{Djouadi}(2008)}]{Djouadi:2005gi}
\bibinfo{author}{\bibfnamefont{A.}~\bibnamefont{Djouadi}},
  \bibinfo{journal}{Phys. Rept.} \textbf{\bibinfo{volume}{457}},
  \bibinfo{pages}{1} (\bibinfo{year}{2008}), \eprint{hep-ph/0503172}.

\bibitem[{\citenamefont{Lester and Summers}(1999)}]{Lester:1999tx}
\bibinfo{author}{\bibfnamefont{C.~G.} \bibnamefont{Lester}} \bibnamefont{and}
  \bibinfo{author}{\bibfnamefont{D.~J.} \bibnamefont{Summers}},
  \bibinfo{journal}{Phys. Lett.} \textbf{\bibinfo{volume}{B463}},
  \bibinfo{pages}{99} (\bibinfo{year}{1999}), \eprint{hep-ph/9906349}.

\bibitem[{\citenamefont{Barr et~al.}(2003)\citenamefont{Barr, Lester, and
  Stephens}}]{Barr:2003rg}
\bibinfo{author}{\bibfnamefont{A.}~\bibnamefont{Barr}},
  \bibinfo{author}{\bibfnamefont{C.}~\bibnamefont{Lester}}, \bibnamefont{and}
  \bibinfo{author}{\bibfnamefont{P.}~\bibnamefont{Stephens}},
  \bibinfo{journal}{J. Phys.} \textbf{\bibinfo{volume}{G29}},
  \bibinfo{pages}{2343} (\bibinfo{year}{2003}), \eprint{hep-ph/0304226}.

\bibitem[{\citenamefont{Kim}(2010)}]{Kim:2009si}
\bibinfo{author}{\bibfnamefont{I.-W.} \bibnamefont{Kim}},
  \bibinfo{journal}{Phys.Rev.Lett.} \textbf{\bibinfo{volume}{104}},
  \bibinfo{pages}{081601} (\bibinfo{year}{2010}), \eprint{0910.1149}.

\bibitem[{\citenamefont{Rujula and Galindo}(2011)}]{Rujula:2011qn}
\bibinfo{author}{\bibfnamefont{A.}~\bibnamefont{Rujula}} \bibnamefont{and}
  \bibinfo{author}{\bibfnamefont{A.}~\bibnamefont{Galindo}},
  \bibinfo{journal}{JHEP} \textbf{\bibinfo{volume}{1108}}, \bibinfo{pages}{023}
  (\bibinfo{year}{2011}), \eprint{1106.0396}.

\bibitem[{\citenamefont{De~Rujula and Galindo}(2012)}]{DeRujula:2012ns}
\bibinfo{author}{\bibfnamefont{A.}~\bibnamefont{De~Rujula}} \bibnamefont{and}
  \bibinfo{author}{\bibfnamefont{A.}~\bibnamefont{Galindo}}
  (\bibinfo{year}{2012}), \eprint{1202.2552}.

\bibitem[{\citenamefont{Barr et~al.}(2009{\natexlab{b}})\citenamefont{Barr,
  Gripaios, and Lester}}]{Barr:2009jv}
\bibinfo{author}{\bibfnamefont{A.~J.} \bibnamefont{Barr}},
  \bibinfo{author}{\bibfnamefont{B.}~\bibnamefont{Gripaios}}, \bibnamefont{and}
  \bibinfo{author}{\bibfnamefont{C.~G.} \bibnamefont{Lester}},
  \bibinfo{journal}{JHEP} \textbf{\bibinfo{volume}{11}}, \bibinfo{pages}{096}
  (\bibinfo{year}{2009}{\natexlab{b}}), \eprint{0908.3779}.

\bibitem[{\citenamefont{van Neerven et~al.}(1982)\citenamefont{van Neerven,
  Vermaseren, and Gaemers}}]{vanNeerven:1982mz}
\bibinfo{author}{\bibfnamefont{W.~L.} \bibnamefont{van Neerven}},
  \bibinfo{author}{\bibfnamefont{J.~A.~M.} \bibnamefont{Vermaseren}},
  \bibnamefont{and} \bibinfo{author}{\bibfnamefont{K.~J.~F.}
  \bibnamefont{Gaemers}} (\bibinfo{year}{1982}),
  \bibinfo{note}{{NIKHEF-H/82-20}}.

\bibitem[{\citenamefont{Arnison et~al.}(1983)}]{Arnison:1983rp}
\bibinfo{author}{\bibfnamefont{G.}~\bibnamefont{Arnison}} \bibnamefont{et~al.}
  (\bibinfo{collaboration}{UA1}), \bibinfo{journal}{Phys. Lett.}
  \textbf{\bibinfo{volume}{B122}}, \bibinfo{pages}{103} (\bibinfo{year}{1983}).

\bibitem[{\citenamefont{Serna}(2008)}]{Serna:2008zk}
\bibinfo{author}{\bibfnamefont{M.}~\bibnamefont{Serna}},
  \bibinfo{journal}{JHEP} \textbf{\bibinfo{volume}{06}}, \bibinfo{pages}{004}
  (\bibinfo{year}{2008}), \eprint{0804.3344}.

\bibitem[{\citenamefont{Cheng and Han}(2008)}]{Cheng:2008hk}
\bibinfo{author}{\bibfnamefont{H.-C.} \bibnamefont{Cheng}} \bibnamefont{and}
  \bibinfo{author}{\bibfnamefont{Z.}~\bibnamefont{Han}},
  \bibinfo{journal}{JHEP} \textbf{\bibinfo{volume}{12}}, \bibinfo{pages}{063}
  (\bibinfo{year}{2008}), \eprint{0810.5178}.

\bibitem[{\citenamefont{Cho et~al.}(2008{\natexlab{a}})\citenamefont{Cho, Choi,
  Kim, and Park}}]{Cho:2007qv}
\bibinfo{author}{\bibfnamefont{W.~S.} \bibnamefont{Cho}},
  \bibinfo{author}{\bibfnamefont{K.}~\bibnamefont{Choi}},
  \bibinfo{author}{\bibfnamefont{Y.~G.} \bibnamefont{Kim}}, \bibnamefont{and}
  \bibinfo{author}{\bibfnamefont{C.~B.} \bibnamefont{Park}},
  \bibinfo{journal}{Phys. Rev. Lett.} \textbf{\bibinfo{volume}{100}},
  \bibinfo{pages}{171801} (\bibinfo{year}{2008}{\natexlab{a}}),
  \eprint{0709.0288}.

\bibitem[{\citenamefont{Gripaios}(2008)}]{Gripaios:2007is}
\bibinfo{author}{\bibfnamefont{B.}~\bibnamefont{Gripaios}},
  \bibinfo{journal}{JHEP} \textbf{\bibinfo{volume}{02}}, \bibinfo{pages}{053}
  (\bibinfo{year}{2008}), \eprint{0709.2740}.

\bibitem[{\citenamefont{Barr et~al.}(2008)\citenamefont{Barr, Gripaios, and
  Lester}}]{Barr:2007hy}
\bibinfo{author}{\bibfnamefont{A.~J.} \bibnamefont{Barr}},
  \bibinfo{author}{\bibfnamefont{B.}~\bibnamefont{Gripaios}}, \bibnamefont{and}
  \bibinfo{author}{\bibfnamefont{C.~G.} \bibnamefont{Lester}},
  \bibinfo{journal}{JHEP} \textbf{\bibinfo{volume}{02}}, \bibinfo{pages}{014}
  (\bibinfo{year}{2008}), \eprint{0711.4008}.

\bibitem[{\citenamefont{Cho et~al.}(2008{\natexlab{b}})\citenamefont{Cho, Choi,
  Kim, and Park}}]{Cho:2007dh}
\bibinfo{author}{\bibfnamefont{W.~S.} \bibnamefont{Cho}},
  \bibinfo{author}{\bibfnamefont{K.}~\bibnamefont{Choi}},
  \bibinfo{author}{\bibfnamefont{Y.~G.} \bibnamefont{Kim}}, \bibnamefont{and}
  \bibinfo{author}{\bibfnamefont{C.~B.} \bibnamefont{Park}},
  \bibinfo{journal}{JHEP} \textbf{\bibinfo{volume}{02}}, \bibinfo{pages}{035}
  (\bibinfo{year}{2008}{\natexlab{b}}), \eprint{0711.4526}.

\bibitem[{\citenamefont{Corcella et~al.}(2002)}]{Corcella:2002jc}
\bibinfo{author}{\bibfnamefont{G.}~\bibnamefont{Corcella}} \bibnamefont{et~al.}
  (\bibinfo{year}{2002}), \eprint{hep-ph/0210213}.

\bibitem[{\citenamefont{Marchesini et~al.}(1992)}]{Marchesini:1991ch}
\bibinfo{author}{\bibfnamefont{G.}~\bibnamefont{Marchesini}}
  \bibnamefont{et~al.}, \bibinfo{journal}{Comput. Phys. Commun.}
  \textbf{\bibinfo{volume}{67}}, \bibinfo{pages}{465} (\bibinfo{year}{1992}).

\bibitem[{her()}]{herwig-bug-fix}
\urlprefix\url{http://projects.hepforge.org/fherwig/trac/ticket/37}.

\end{thebibliography}
\end{document}